\documentstyle[12pt,aps]{revtex}

\begin{document}
\draft
\title{Extracting energies from the vacuum}
\author{She-Sheng Xue
}
\address{
ICRA, INFN  and
Physics Department, University of Rome ``La Sapienza", 00185 Rome, Italy
}

\date{August, 2000}


\maketitle

\centerline{xue@icra.it}

\begin{abstract}

We present and study a possible mechanism of extracting energies from the vacuum by external classical fields. Taking a constant magnetic field as an example, we discuss why and how the vacuum energy can be released in the context of quantum field theories. In addition, we give a theoretical computation showing how much vacuum energies can be released. The possibilities of experimentally detecting such a vacuum-energy releasing are discussed.

\end{abstract}


\narrowtext

The vacuum has a very rich physical content in the context of relativistic quantum field theories. It consists of extremely large number of virtual particles and anti-particles. The quantum fluctuations of the vacuum are creations and annihilations of these virtual particles and antiparticles in all possible energy-range. As a consequence of the quantum fluctuations, the vacuum energy in a  volume $V$ of the three-dimensional space is given by 
\begin{equation}
{\cal E} =sV\sum_k \epsilon(k),
\label{ve}
\end{equation}
where the $\epsilon(k)$ is the energy-spectrum of virtual particles, the summation is over all possible momentum-states of quantum-field fluctuations and the factor $s$ is $1/2$ for electromagnetic fields and $1$ for fermionic fields. The vacuum energy (\ref{ve}) is related to the transition amplitude $\langle 0 | 0 \rangle$ from the vacuum to the vacuum, which is represented by close loops for virtual particles  in Feynman diagrams. Up to the fundamental Planck scale $\Lambda_p\sim 10^{19}$GeV, the vacuum-energy $|{\cal E}|\sim V\Lambda_p^4$, which is a tremendous amount of energies for a macroscopic volume $V$. According to quantum field theories for free and massive particles, the energy-spectrum of virtual particles in the vacuum is given by
\begin{equation}
\epsilon(k)=\pm\sqrt{(ck_x)^2+(ck_y)^2+(ck_z)^2+m^2c^4},
\label{s}
\end{equation}
where $m$ is the mass of particles, the sign ``+'' is for electromagnetic  fields and ``-'' for fermionic fields. In the case of electromagnetic fields ($m=0$), the virtual photon and its anti-particle are self-conjugated and all virtual photons can be accommodated in the same energy-momentum state. The positive vacuum-energy (\ref{ve}) is due to the quantum fluctuations of these virtual photons in all possible energy-range. In the case of fermionic fields, a pair of virtual fermion and anti-fermion occupy a energy-momentum state and all negative energy states of quantum fermion fields down to the negative energy-level of the Planck scale ($-\Lambda_p$) have been fully filled by the pairs of virtual fermions and anti-fermions. The virtual fermion in the negative energy state travels backward in time indicating a virtual anti-fermion in the positive energy state travels forward in time. The negative vacuum-energy (\ref{ve}) is due to the quantum fluctuations of the creations (annihilations) of virtual fermions and anti-fermions by (to) virtual photons in all possible energy-range. The pairs of virtual fermions and anti-fermions are created and annihilated in the time scale $\sim\hbar/mc^2$ and at the distance scale $\sim\hbar/mc$.

In the description of renormalizable and perturbative quantum field theories, the vacuum state is a ground state consisting of virtual particles with the energy-spectrum (\ref{s}), all negative energy states are fully filled and real particles are excitation quanta upon this ground state. This description is validated, provided the amplitude of quantum fields and the strength of interactions between quantum fields are small upon such a vacuum state. In this description, the vacuum energy (\ref{ve}) is dropped and set to be zero by the normal ordering of creation and annihilation operators, due to the absolute value of the physical energy only determined up to a constant and the quantum fluctuation of the vacuum impacting on real particles is treated by the renormalization of theories. The descriptions of renormalizable and perturbative quantum field theories have been extremely successful. As an example, the Lamb-shift\cite{lamb} effect and electric charge renormalization indeed exhibit the highly non-trivial structure of the quantum electromagnetic dynamics (QED) and its vacuum (ground) state. 

However, as shown by the Casimir effect\cite{casimir} that was experimentally evidenced\cite{exp}, the positive vacuum-energy (\ref{ve}) of the vacuum state is not just a trivial constant, when the quantum fluctuations of virtual photons of the vacuum state are confined within a finite volume by boundary conditions.
The Casimir effect can be physically understood as the following: the continuous energy-spectrum (\ref{s}) of electromagnetic fields is modified by boundary conditions to be discrete one, the ``new'' vacuum energy computed by the modified energy-spectrum in a given finite volume is smaller than the ``old'' vacuum energy computed by the energy-spectrum (\ref{s}) in the same volume. As a result, the vacuum has to quantum-mechanically fluctuate from the ``old'' vacuum state to the ``new'' vacuum state. This leads to an attractive force observed in the Casimir effect. This is in fact a phenomenon of extracting energies from the vacuum. On the contrary, if the ``new'' vacuum energy computed by the modified energy-spectrum is larger than the ``old'' vacuum energy computed by the energy-spectrum (\ref{s}), an external force must be introduced to against a repulsive force and store the work done by the external force into the vacuum energy.

In this letter, instead of modifying the energy-spectrum (\ref{s}) of electromagnetic fields by boundary conditions, we attempt to study the vacuum-energy releasing  by modifying the negative energy-spectrum for virtual fermions with external classical fields and to find any possibly observable effects of the vacuum-energy releasing due to such a modification. 

We first take the case of a constant magnetic field as an example to discuss and illustrate the reasons for the 
vacuum-energy releasing. The negative energy-spectrum (\ref{s}) for virtual fermions is not degenerate. The vacuum state is made by virtual fermions fully filling this negative energy-spectrum. We apply the external constant magnetic field $H$ onto such a vacuum state along the z-axis. This constant magnetic field is confined within a finite space of a volume $V=L_x\cdot L_y\cdot L_z$. As well known as the Landau levels\cite{landau}, the negative energy-spectrum of virtual charged fermions is given by
\begin{equation}
\epsilon(k_z,n,\sigma)=-\sqrt{(ck_z)^2+m^2c^4+\hbar c|e|H(2n+1)-\hbar c eH\sigma},\hskip0.3cm n=0,1,2,3,\cdot\cdot\cdot ,
\label{sh}
\end{equation}   
where $e$ and $\sigma=\pm 1$ are fermion's charge and helicity. This negative energy-spectrum is degenerate in the phase space of ($k_x,k_y$) and the degeneracy is $|e|HS/(2\pi\hbar c)$, where the area $S=L_x\cdot L_y$. 

The negative energy-spectrum is changed from (\ref{s}) to (\ref{sh}) due to the external magnetic field $H$. If the vacuum energy of the ``new'' vacuum state made by virtual fermions fully filling the negative energy-spectrum (\ref{sh}) is smaller than  that of the ``old'' vacuum state made by virtual fermions fully filling the negative energy-spectrum (\ref{s}). The ``old'' vacuum state must quantum-mechanically  fluctuate to the ``new'' vacuum state. The difference of vacuum energies between two vacuum states must be released. In order to verify this and possibly observable effects, we are bound to compute the energetic  difference between two vacuum states corresponding to $H=0$ and $H\not= 0$. 

The total energies of the ``old'' vacuum state and ``new'' vacuum state are respectively given by,
\begin{equation}
{\cal E}_o=-4\Big({V\over(2\pi\hbar c)^3}\Big)\int d(c\vec k)|\epsilon(k)|,
\label{ove}
\end{equation}
and     
\begin{equation}
{\cal E}_n=-\Big({|e|HS\over2\pi\hbar c}\Big)\Big({ L_z\over2\pi\hbar}\Big)\int dk_z\sum_{n,\sigma}|\epsilon(k_z,n,\sigma)|.
\label{nve}
\end{equation}
Both vacuum energies are divergent up to the Planck scale $-V\cdot\Lambda_p^4$. The difference of vacuum energies between ``new'' and ``old'' vacuum states can be computed by the approaches of the dimensional regularization\cite{thooft} and $\xi$-function regularization\cite{db}. In eq.(\ref{ove}), analytically continuing the dimension of the momentum integration from $3$ to $3+\epsilon$, where $\epsilon$ is a small complex parameter, we have
\begin{equation}
{\cal E}_o=\Big({V\pi\over(2\pi\hbar c)^3}\Big)(mc^2)^4\Gamma(-{\epsilon\over2}),
\label{ove1}
\end{equation}
where $\Gamma(x)$ is the Gamma-function. Analogously, in eq.(\ref{nve}), analytically continuing the dimension of the momentum integration from $1$ to $1+\epsilon$, we have,
\begin{equation}
{\cal E}_n =-\Big({|e|HV\over4\pi^2\hbar^2 c^2}\Big)\Gamma(-{\epsilon\over2})\sum_{n,\sigma}\Big[m^2c^4+\hbar c|e|H(2n+1)-\hbar c eH\sigma\Big]^z,
\label{dnve}
\end{equation}
where $z=1+\epsilon/2$. Summing over helicity states $\sigma=\pm 1$, we obtain 
\begin{eqnarray}
{\cal E}_n &=&-\Big({|e|HV\over4\pi^2\hbar^2 c^2}\Big)\Gamma(-{\epsilon\over2})\sum_n\left[\Big(m^2c^4+2\hbar c|e|Hn\Big)^z+ \Big(m^2c^4+2\hbar c|e|H(n+1)\Big)^z\right],\nonumber\\
& = & -\Big({2|e|^2H^2V\over4\pi^2\hbar c}\Big)\Gamma(-{\epsilon\over2})\left[\xi(-z,q)+\xi(-z,q+1)\right],\hskip0.2cm q={m^2c^4\over 2\hbar c |e|H},
\label{dnve1}
\end{eqnarray}
where $\xi(z,q)$-function is given by\cite{gr},
\begin{equation}
\xi(z,q)=\sum_{n=0}{1\over (n+q)^z},\hskip0.3cm\xi(-1,q)=-{B_2(q)\over2},\hskip0.3cm B_2(q)=q^2-q+{1\over6}.
\label{xi}
\end{equation}
The analytic continuation ``$z$'' has simply discarded the appropriate divergent terms and the continuation back to ``$z=1$'' ($\epsilon\rightarrow 0$) yields,
\begin{equation}
{\cal E}_n =V\Gamma(-{\epsilon\over2})\left[{|e|^2H^2\over3(4\pi^2\hbar c)}+ {\pi (m^2c^4)^2\over(2\pi\hbar c)^3}\right].
\label{dnve2}
\end{equation}
As a result, the energetic  difference of ``new'' and ``old'' vacuum states is,
\begin{equation}
\Delta{\cal E}={\cal E}_n - {\cal E}_o =\Gamma(-{\epsilon\over2})\left[{|e|^2H^2V\over3(4\pi^2\hbar c)}\right].
\label{delta1}
\end{equation}
We find that in eq.(\ref{dnve2})the term depending on the fermion mass is completely canceled by eq.(\ref{ove1}), as it should be. For $\epsilon\rightarrow 0$, the Gamma-function $\Gamma(-\epsilon/2)=-(2/\epsilon + {\rm const.})$, where the constant is an uninteresting combination of $\pi,\gamma$(Euler constant), etc. On the basis of the charge renormalization of the QED, we renormalize the charge $|e_r|=\sqrt{Z_3}|e|$ by the renormalization constant $Z_3=(2/\epsilon + {\rm const.})$ and we obatin,
\begin{equation}
\Delta{\cal E}=-\sum_f(Q_f^2){e_r^2H^2V\over3(4\pi^2\hbar c)}=-8{\alpha \over3\pi}H^2V, \hskip0.5cm \alpha={e_r^2\over4\pi\hbar c},
\label{delta2}
\end{equation}
where the fine structure constant $\alpha=1/137$ and $\sum_f(Q_f^2)=8$ for all charged fermions in the standard model of particle physics. Eq.(\ref{delta2}) is about one percent of the total energy deposited by the external magnetic field $H$.

In principle, since $\Delta{\cal E}$ is negative, indicating that the energy of the ``new'' vacuum for $H\not=0$ is smaller than that of the ``old'' vacuum for $H=0$, the vacuum must release the energy (\ref{delta2}) when the external magnetic field is applied upon it.  Let us try to understand why and how this could physically occur. This can possibly occur for three reasons that (i) in a finite volume $V$, the total number of fermionic states in the vacua of negative energy-spectrum (\ref{s}) and (\ref{sh}) is finite for the finite momentum-cutoff at the Planck scale $\Lambda_p$ and all these fermionic states of negative energy levels from $-\Lambda_p$ to $-mc^2$ are fully filled; (ii) the negative energy-spectrum (\ref{s}) is not degenerate, while the negative energy-spectrum (\ref{sh}) is degenerate, and the total
numbers of fermionic states of both cases are the same as indicated in (i); (iii) on the basis of quantum fluctuations toward the lowest energy-state and the Pauli principle, , when the external magnetic field is applied upon the vacuum, the vacuum reorganize itself by fully filling all fermion states according to the degenerate negative energy-spectrum (\ref{sh}), instead of the non-degenerate one (\ref{s}). As a consequence, the vacuum makes its total energy lower and releases energies. 
As an analogy, the vacuum with the negative energy-spectrum (\ref{s}) can be described as if a $N$-flours building, two rooms each flour and all rooms occupied by guests; while the vacuum with the negative energy-spectrum (\ref{sh}) a $M$-flours $(M<N)$ building, $2N/M$ rooms each flour and all rooms occupied by guests. The total numbers of rooms of two buildings are the same. Due to an external force, the $N$-flours building collapses to the $M$-flours building and the ``potential energy'' should be released. 

If the vacuum energy (\ref{delta2}) is released, the phenomena and effects that could occur are following.  (i) The vacuum acts as a paramagnetic medium that effectively screens the strength of the external magnetic field $H$ to a smaller value $H'<H$ for the total energy-density being
\begin{equation}
{1\over2}H'^2={1\over2}H^2-8{\alpha \over3\pi}H^2;\hskip0.5cm H'=H\sqrt{1-{16\over3\pi}\alpha}.
\label{hed}
\end{equation}
(ii) Photons are spontaneously emitted analogously to the spontaneous emission for electrons at high-energy levels decaying to low-energy levels in the atomic physics. (iii) This strong vacuum-energy fluctuation could lead to the emission of neutrino and anti-neutrino pairs from the vacuum, since they are almost massless. Such a vacuum-energy releasing phenomena can be certainly checked by appropriate experiments of precisely measuring the magnetic field strength and detecting any photons emitted when the magnetic field is turned on. 

In practice, on the other hand, how the phenomena of the vacuum-energy releasing can be realized and observed in reality. In the theoretical analysis given, we actually assume that the transition of the vacuum states from the negative energy-spectrum (\ref{s}) to the negative energy-spectrum (\ref{sh}) is instantaneous. In reality, there is a time scale $\Delta \tau$ for building up the magnetic field to its maximum value. The maximum volume $(\Delta\tau\cdot c)^3$, in which vacuum states at different points of the space-time are causally-correlated, can be much larger than the volume $V$ in which the constant magnetic field is created. In addition, we know that the relaxation time $\Delta t$ of quantum fluctuations of vacuum states should be the order of $\hbar/m_ec^2\sim 10^{-20}$sec. With this very short relaxation time, all causally-correlated vacuum states at different points of the space-time can rapidly decay to the lower energy states before the maximum value of the external magnetic field is reached. In this case, we should not expect to detect appropriate photon emissions, corresponding to the total vacuum-energy releasing given in eq.(\ref{delta2}), only from the volume $V$ where the constant magnetic field is created, since photons can be spontaneously emitted from the whole volume $\sim(\Delta\tau\cdot c)^3$ where causally-correlated vacuum states are. If we can create the magnetic field in a short enough time, we should expect to detect photon emissions and other phenomena discussed in the previous paragraph. 

In contraction to the Casimir effect, two large parallel perfectly conducting plates of sizes $L^2$ at a distance $a$ ($L\gg a$) separate causally-correlation of the quantum fluctuations of virtual photons inside the volume $aL^2$ between two plates from the quantum fluctuations of virtual photons outside of two plates. Since the creations and annihilations of virtual fermions and antifermions is related to the annihilation and creation of virtual photons, we assume the causally-correlation between vacuum states at two different points of the space-time is mediated by virtual photons. Inspired by the Casimir effect, we should use perfectly conducting box to isolate the volume $V$, where the magnetic field is created, from the space-time outside the box. The photon monitor and instrument of measuring magnetic field strength should be properly installed inside the box to detect possible effect of the vacuum-energy releasing.

It is worthwhile to mention that with respect to an observer at infinite, such effect (with photon and neutrino emissions) for a very large strength of magnetic fields in astrophysics (e.g., neutron starts) could be probably observed. Given the size of a neutron start of the order $10^6$cm and the strength of magnetic fields of the order of $10^{13}-10^{15}G$, we can estimate the maximum total vacuum-energy releasing is of the order of $10^{42}-10^{46}$erg from eq.(\ref{delta2}).    

As a comparison, we consider the case of an external electric field by a simple example. For a spherical conductor with the radius $R_\circ$ and charge $Q_\circ<0$,
we have the electric potential $U$ and field $E$, $U=U_\circ=Q_\circ/R_\circ$ and $E=0$ for $r\le R_\circ$;  $U=Q_\circ/r$ and $E=Q_\circ/r^2$ for $r > R_\circ$. We postulate the electric potential energy $eU_\circ>mc^2$ and the  field $E$ at $R_\circ$ is less than the critical value to create electron-positron pairs. It is rather complicate to find the exact negative energy-spectrum in such an attractive potential. To obtain a preliminary result, we may approximate this potential by a spherical potential well $U=U_\circ, r\le R_\circ$; $U=0, r>R_\circ$. The first negative discrete energy-level ${\cal E}_1$ of an electron of charge $e<0$ is given by\cite{landau2}
\begin{equation}
{\cal E}_1 = -{\pi^2\over16}U_\circ^{\rm min}\Delta^2, \hskip0.3cm U_\circ^{\rm min}={(\pi\hbar c)^2\over8mc^2R^2_\circ},\nonumber\\
\hskip0.3cm \Delta^2 = \left({eU_\circ\over U_\circ^{\rm min}}-1\right)^2
\label{e1}
\end{equation}
for small $\Delta$, where $U_\circ^{\rm min}$ is the minimum depth of the potential well for appearing the first negative discrete energy-level. For the macroscopic value of the radius $R_\circ\gg \hbar/(m_ec)$, $U_\circ^{\rm min}$ and ${\cal E}_1$ are extremely small. Based on this approximate negative energy-spectrum, we can estimate the difference between the total vacuum energies ${\cal E}_n $ with and ${\cal E}_o$ without the presence of the external electric potential $U_\circ$,  
\begin{equation}
\Delta{\cal E}={\cal E}_n - {\cal E}_o\simeq {\cal E}_1<0.
\label{delta3}
\end{equation}
This simply shows that the negative energy-spectrum of the vacuum is shifted downward from $-mc^2$ to $-mc^2+{\cal E}_1$. The possible vacuum-energy releasing $|{\cal E}_1|$ in this case is extremely small for the radius $R_\circ\gg \hbar/(mc)$. The result of the similar idea applying to the case of an external gravitational field will be presented in another paper\cite{gravity}.

In this letter, we propose an idea of extracting vacuum-energy from the vacuum by external classical-fields. This idea is originated from (i) the fermionic structure of the vacuum owing to relativistic quantum field theories and the Pauli principle; (ii) external classical-fields possibly modifying the fermionic structure of the vacuum; (iii) the quantum fluctuations of virtual particles in the vacuum must lead the vacuum to the lowest energy-state. Taking a particular example of the constant magnetic field applied upon the vacuum, we illustrate this idea in very detail and give an explicit computation showing the vacuum-energy releasing is about one percent of the total energy stored by the external magnetic field. In addition, we propose a possible experiment to detect possible phenomena and effects due to the vacuum-energy releasing.    
In today's laboratory, the magnetic field strength has been reached up to (greater than) 10 T and large stored energy up to (larger than) tens or hundreds of MJ \cite{magnet}. We expect an appropriate experiment to verify the phenomena and effects of the vacuum-energy releasing induced by the external magnetic field discussed in this letter. This is important for the understanding of the fermionic structure of the vacuum of relativistic quantum field theories and any possibly prospective applications\cite{application}.



\begin{references}
\bibitem{application}
The right to use this idea (as an intelligent product) for any applications is preserved by the author.
\end{references}
\end{document}